\documentclass[11pt, a4paper]{article}
\pdfoutput=1

\usepackage{jinstpub}
\usepackage{natbib}

\usepackage{graphicx}
\usepackage{dcolumn}
\usepackage{bm}
\usepackage{siunitx}
\usepackage[utf8]{inputenc}
\usepackage[USenglish]{babel}

\usepackage{rotating}
\usepackage{booktabs}
\usepackage{hyperref}
\usepackage{subcaption}

\DeclareSIUnit\meter{m}

\begin{document}

\title{Beam-based alignment at the Cooler Synchrotron COSY as a prerequisite for an electric dipole moment measurement}

\collaboration{\includegraphics[height = 17mm]{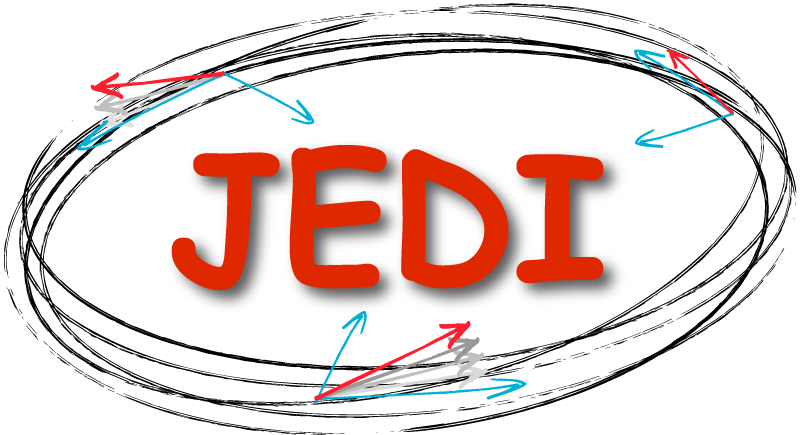}\\[6pt]JEDI collaboration}

\author[a,b,1]{T.~Wagner,}
\note{corresponding author}
\author[b]{A.~Nass,}
\author[a,b,c]{J.~Pretz,}
\author[a,b]{F.~Abusaif,}
\author[d]{A.~Aggarwal,}
\author[a,b]{A.~Andres,}
\author[b]{I.~Bekman,}
\author[e]{N.~Canale,}
\author[f]{I.~Ciepal,}
\author[e]{G.~Ciullo,}
\author[b]{F.~Dahmen,}
\author[e,g]{S.~Dymov,}
\author[b]{C.~Ehrlich,}
\author[b]{R.~Gebel,}
\author[a,b]{K.~Grigoryev,}
\author[b]{D.~Grzonka,}
\author[b]{V.~Hejny,}
\author[b]{J.~Hetzel,}
\author[b]{A.~Kacharava,}
\author[b]{V.~Kamerdzhiev,}
\author[d]{S.~Karanth,}
\author[b]{I.~Keshelashvili,}
\author[e]{A.~Kononov,}
\author[g]{A.~Kulikov,}
\author[a]{K.~Laihem,}
\author[b,c]{A.~Lehrach,}
\author[e]{P.~Lenisa,}
\author[h]{N.~Lomidze,}
\author[d]{A.~Magiera,}
\author[h]{D.~Mchedlishvili,}
\author[a,b]{F.~M\"uller,}
\author[i,j]{N.N.~Nikolaev,}
\author[e]{A.~Pesce,}
\author[a,b]{V.~Poncza,}
\author[b]{F.~Rathmann,}
\author[b]{M.~Retzlaff,}
\author[b]{A.~Saleev,}
\author[b]{M.~Schm\"uhl,}
\author[h]{D.~Shergelashvili,}
\author[b,g]{V.~Shmakova,}
\author[k]{J.~Slim,}
\author[a]{A.~Stahl,}
\author[l]{E.~Stephenson,}
\author[b,c]{H.~Str\"oher,}
\author[h]{M.~Tabidze,}
\author[m]{G.~Tagliente,}
\author[n]{R.~Talman,}
\author[g,o,p]{Yu.~Uzikov,}
\author[b]{Yu.~Valdau}
\author[d]{and A.~Wro\'{n}ska}

\affiliation[a]{III. Physikalisches Institut B, RWTH Aachen University, 52056 Aachen, Germany}
\affiliation[b]{Institut f\"ur Kernphysik, Forschungszentrum J\"ulich, 52425 J\"ulich, Germany}
\affiliation[c]{JARA--FAME (Forces and Matter Experiments), Forschungszentrum J\"ulich and RWTH Aachen University, Germany}
\affiliation[d]{Marian Smoluchowski Institute of Physics, Jagiellonian University, 30348 Cracow, Poland}
\affiliation[e]{University of Ferrara and INFN, 44100 Ferrara, Italy}
\affiliation[f]{Institute of Nuclear Physics, Polish Academy of Sciences, 31342 Crakow, Poland}
\affiliation[g]{Laboratory of Nuclear Problems, Joint Institute for Nuclear Research, 141980 Dubna, Russia}
\affiliation[h]{High-Energy Physics Institute, Tbilisi State University, 0186 Tbilisi, Georgia}
\affiliation[i]{L.D. Landau Institute for Theoretical Physics, 142432 Chernogolovka, Russia}
\affiliation[j]{Moscow Institute for Physics and Technology, 141700 Dolgoprudny, Russia}
\affiliation[k]{Institut f\"ur Hochfrequenztechnik, RWTH Aachen University, 52056 Aachen, Germany}
\affiliation[l]{Indiana University Center for Spacetime Symmetries, Bloomington,  Indiana 47405, USA}
\affiliation[m]{Istituto Nazionale di Fisica Nucleare, 70125 Bari, Italy}
\affiliation[n]{Cornell University, Ithaca,  New York 14850, USA}
\affiliation[o]{Dubna State University, 141980 Dubna, Russia}
\affiliation[p]{Department of Physics, M.V. Lomonosov Moscow State University, 119991 Moscow, Russia}

\emailAdd{t.wagner@fz-juelich.de}

\abstract{
The Jülich Electric Dipole moment Investigation (JEDI) collaboration aims at a direct measurement of the Electric Dipole Moment (EDM) of protons and deuterons using a storage ring.
The measurement is based on a polarization measurement.
In order to reach highest accuracy, one has to know the exact trajectory through the magnets, especially the quadrupoles, to avoid the influence of magnetic fields on the polarization vector.
In this paper, the development of a beam-based alignment technique is described that was developed and implemented at the COoler SYnchrotron (COSY) at Forschungszentrum Jülich.
Well aligned quadrupoles permit one to absolutely calibrate the Beam Position Monitors (BPMs).
The method is based on the fact that a particle beam, which does not pass through the center of a quadrupole, experiences a deflection.
The precision reached by the method is approximately \SI{40}{\micro\metre}. Some consequences for the design 
of a new high precision storage ring for EDM mesasurements are discussed.
}


\arxivnumber{2009.02058}

\maketitle
\flushbottom

\section{Introduction and Motivation}
The observed matter-antimatter asymmetry in the universe cannot be explained by the Standard Model of particle physics and cosmology alone.
Additional CP violating mechanisms beyond the already known effects are needed~\cite{Sakharov:1967dj}.
Evidence for additional CP violating effects is accessible from a measurement of permanent Electric Dipole Moments (EDMs) of subatomic particles.
The EDMs violate both parity and time reversal symmetry, and also violate CP symmetry if the CPT-theorem holds.
However, the EDMs predicted by the Standard Model are orders of magnitudes too small to explain the dominance of matter over antimatter in the universe.
The discovery of a large EDM would hint towards physics beyond the Standard Model and contribute to the explanation for the dominance of matter over antimatter in the universe.

\begin{figure*}
    \centering
    \includegraphics[width = \textwidth]{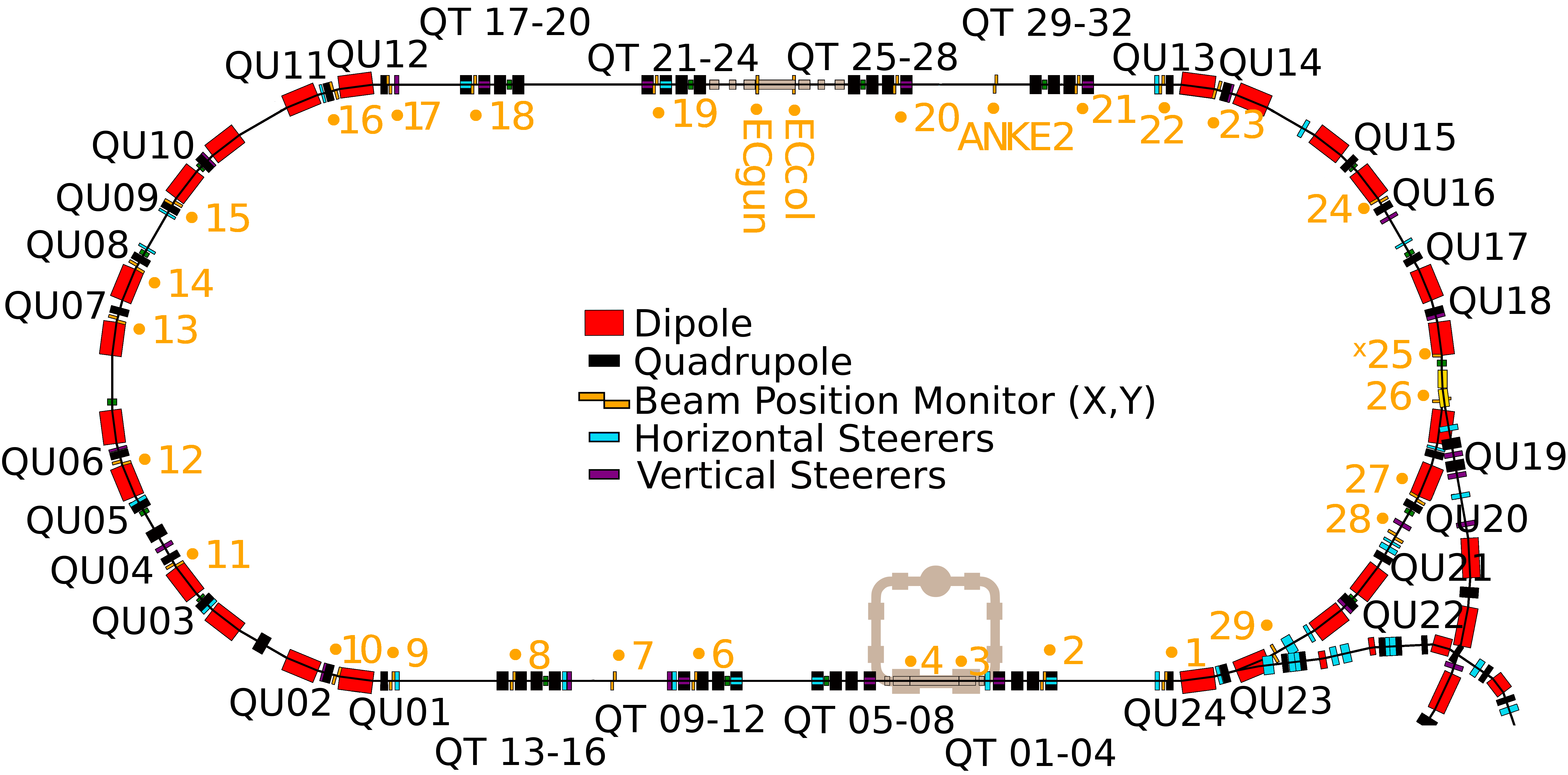}
    \caption{Sketch of COSY~\cite{MAIER1997395} with labeled quadrupoles and BPMs along the ring. The black elements represent the quadrupoles and the yellow elements the BPMs. The quadrupoles in the straight sections are called "QT", whereas the quadrupoles in the arcs are called "QU", which is used to distinguish them due to a different width of the magnets. The dipoles are shown in red and horizontal and vertical steerers in light blue and purple, respectively.}
    \label{fig:COSY_sketch}
\end{figure*}

The observation of EDMs of subatomic particles is possible by observing their interaction with electric fields. For neutral particles (e.g., the neutron~\cite{Abel:2020gbr}) this can be done in small volumes. 
Because of their acceleration in electric fields, for charged particles this constitutes a more difficult task. The Jülich Electric Dipole moment Investigation (JEDI) Collaboration aims to measure the EDM of the proton and the deuteron in a storage ring.
The control of systematic errors is of paramount importance, making the design of a dedicated EDM storage ring, and in particular the alignment of the ring elements~\cite{Abusaif:2019gry} and the correction of the closed orbit~\cite{Rosenthal:2016zbf} a very demanding task.
At the Cooler Synchrotron (COSY) (see figure~\ref{fig:COSY_sketch}) of Forschungszentrum Jülich, a first storage ring EDM measurement is presently being carried out for deuterons~\cite{Lenisa:2020jxb} and is being planned for protons.
In order to improve the precision of the machine, a beam-based alignment method is applied to align the magnetic centers of the quadrupole magnets and the BPMs.
This beam-based alignment method has been applied at electron~\cite{Portman:1995py, Ogur:2019rsn, Jena:2015uva} and hadron~\cite{Niedziela:2005qr, Hayashi:2010zz} machines before.
This procedure requires the quadrupoles to be mechanically aligned.
This was achieved by a surveying procedure\footnote{Vermessungsbüro Stollenwerk \& Burghof, 50126 Bergheim, Germany} to a precision of \SI{200}{\micro\metre}. The detailed alignment data are listed in table~\ref{tab:quad_alignment_straight} and table~\ref{tab:quad_alignment_arc} in the appendix.
Several tests were performed to determine the effect of a full beam-based alignment survey of all quadrupoles in the accelerator could have. The first two tests~\cite{Wagner:2018qjg, Wagner:2019xul} showed that the offset between the quadrupoles and BPMs amounts to several \si{\milli\metre}. Thus, the beam-based alignment measurement has to be performed for all quadrupoles in the accelerator in order to have all BPMs properly calibrated.

The full measurement campaign for the beam-based alignment has been performed at COSY for all 56 quadrupoles in the ring. 
With the use of the 31 BPMs it was possible to determine the center of the 56 quadrupoles in COSY and with that calculate the offset between the BPMs and quadrupoles in order to get a better offset calibration of the BPMs.
In addition this measurement also allowed to check the alignment of the quadrupoles in the straight sections of the accelerator with respect to each other, where it was found out that some magnets are not on axis with the other quadrupoles (in contrast to the alignment survey results).

This paper is organized as follows. In section~\ref{sec:theory} a short introduction is given, in which the method is described. Section~\ref{sec:cosy} describes the measurement at COSY and the analysis of the data followed by section~\ref{sec:res} discussing the results.

\section{Theoretical description of beam-based alignment}\label{sec:theory}
In order to determine whether a particle beam passes through the center of a quadrupole,  one can use the effect that an off-center beam experiences a dipole component leading to a kick of the particle beam.
By varying the quadrupole strength, one simultaneously varies the magnitude of the dipole component of the field, which the off-center beam experiences.
This results in a measurable closed-orbit change that depends on the offset of the beam inside the quadrupole, whose strength was varied.
The orbit change in one plane (here horizontal, x) can be described by~\cite{Portman:1995py}
\begin{equation}
\Delta x(s) = \frac{\Delta k\cdot x(s_0)\ell}{B\rho} \frac{1}{1-k\frac{\ell\beta(s_0)}{2B\rho\tan \pi \nu}} \frac{\sqrt{\beta(s)}\sqrt{\beta(s_0)}}{2\sin\pi\nu}\cos[\phi(s)-\phi(s_0)-\pi\nu],
\label{eq:orbitChange}
\end{equation}
where the parameters are explained in table~\ref{tab:parFormula}.

\begin{table}[!htbp]
\caption{Explanation of the parameters in (\ref{eq:orbitChange})}
\label{tab:parFormula}
\centering
\bgroup
\def\arraystretch{1.15}
\begin{tabular}{cp{6.5cm}}
\toprule
Parameter & Meaning \\\midrule
$\Delta x$ & Orbit change\\
$s$ & Measurement position\\
$s_0$ & Position of quadrupole\\
$\Delta k$ & Change in quadrupole strength\\
$x(s_0)$ & Position of the beam with respect to the magnetic center of the quadrupole\\
$\ell$ & Length of quadrupole\\
$B\rho = \frac{p}{q}$ & Magnetic rigidity of the beam\\
$k$ & Quadrupole strength\\
$\beta$ & Beta function\\
$\nu$ & Betatron tune\\
$\phi$ & Betatron phase\\
\bottomrule
\end{tabular}
\egroup
\end{table}

From (\ref{eq:orbitChange}) one can see that the orbit change $\Delta x(s)$ at a given position $s$ is proportional to the beam position inside the quadrupole $x(s_0)$.
As not all parameters in (\ref{eq:orbitChange}) are perfectly known all along the accelerator, the proportionality is quite useful.
This permits one to construct a merit function to extract the optimal position of the beam inside the quadrupole from the measured data. 
The merit function that was used for this measurement is

\begin{eqnarray}
f(x(s_0), y(s_0)) &=& \frac{1}{N_{\text{BPM}}} \sum^{N_{\text{BPM}}}_{i = 1} \left[(\Delta x_i)^2 + (\Delta y_i)^2\right]\text{, with}\label{eq:meritFunction}\\
\Delta x_i &=& x_i(x(s_0); +\Delta k) - x_i(x(s_0); -\Delta k),\nonumber\\
\Delta y_i &=& y_i(y(s_0); +\Delta k) - y_i(y(s_0); -\Delta k).\nonumber
\end{eqnarray}

In order to determine the merit function $f(x(s_0), y(s_0))$, one has to take two measurements for each beam position inside the quadrupole, one measurement with slightly increased $(+\Delta k)$ quadrupole strength and another one with slightly reduced $(-\Delta k)$ quadrupole strength.
The differences of the beam positions $x_i$ and $y_i$ at the $i$-th beam position monitor are summed up in quadrature for all beam position monitors (see (\ref{eq:meritFunction})).
It is easy to conclude that the merit function is proportional to the offset of the beam inside the quadrupole squared $f \propto \sum_i (\Delta x_i)^2 + (\Delta y_i)^2 \propto a(x(s_0))^2 + b(y(s_0))^2$, where the factors $a$ and $b$ are introduced because the sensitivity to the strength change of the quadrupole is different in horizontal and vertical direction. 
The shape of the merit function is a paraboloid and by finding its minimum one can determine the optimal position of the beam inside the quadrupole.

\section{Measurements at COSY}\label{sec:cosy}
\subsection{Hardware upgrades at COSY}
In order to perform the measurement each quadrupole strength has to be modified individually. In the powering scheme of COSY though, four quadrupoles are powered by one main power supply.
Some quadrupoles are equipped with back-leg windings, which allows for an individual control. A first beam-based alignment measurement has been performed with those 12 quadrupoles already~\cite{Wagner:2019xul}.

As not all quadrupoles could be equipped with back-leg windings, a new solution was found. 
A smaller floating power supply was added in parallel to one quadrupole magnet in order to add or bypass some of the current. The devices chosen are source-sink power supplies\footnote{Höcherl \& Hackl GmbH \url{https://www.hoecherl-hackl.com/}} and are ideally suited for the beam-based alignment. 
For cost reasons, it was not possible to acquire 56 of these power supplies. Instead it was decided to purchase only four of them and connect them as needed during the beam time. For this purpose, connectors were mounted on each quadrupole, where the mobile power supplies can be plugged in.

The communication with the power supplies was realized with serial communication over Ethernet in order to dynamically control them during the measurement. As an additional safety aspect the power supplies were disconnected during the acceleration of the beam not to interfere with the ramping of the quadrupoles and they were later connected with a relay after the acceleration.

With this hardware upgrade and the existing system of 31 BPMs the measurements described in the next section could be performed.

\subsection{Measurement procedure}

As a first step a rough offset calibration of the BPMs was performed where only the quadrupoles, which have a BPM next to them, were measured for the offset calibration of the BPMs. This first offset calibration was applied and then the measurement for all the quadrupoles was performed as a second step. This first offset calibration was done in order to know the approximate optimal position inside the quadrupoles and use a scan with a better resolution and smaller range afterwards.

\begin{figure}[!tbp]
\centering
\includegraphics[width=0.7\textwidth]{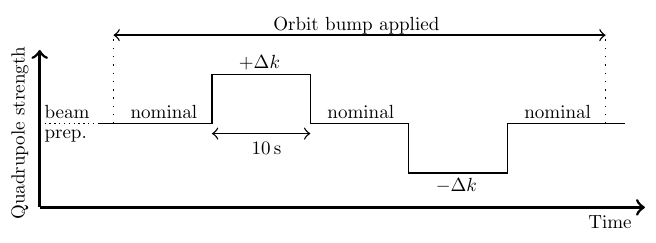}
\caption{Pattern of strength change of the quadrupole during the measurement. In the beginning of the cycle the beam is prepared, i.e. accelerated, bunched and positioned as desired with the use of an orbit bump. Then the quadrupole strength is left at the nominal strength to have a reference point to check if something changes during the measurement time due to outside effects. Next the quadrupole strength is increased by $+\Delta k$ and afterwards again set to nominal strength. Then the quadrupole strength is set to nominal strength $-\Delta k$ and afterwards set to nominal strength again. This is a complete pattern leading to one data point for the merit function.}
\label{fig:measurement_pattern}
\end{figure}

The measurement procedure scans multiple different beam positions inside the quadrupole to find the optimal position, where a strength change does not steer the beam.
For the measurement the beam was prepared and then the additional power supplies were connected with the help of a relay.
Then during the cycle the beam was moved to different positions inside the quadrupoles with the help of nearby steerers using a local orbit bump.
It was made sure that the orbit bump was localized, but this requirement is not needed to be very strict, as only changes of the orbit between the $\pm\Delta k$ quadrupole strength changes are needed to compute the merit function \eqref{eq:meritFunction}, during which the bump was always applied.
This orbit bump was kept while manipulating the quadrupole strength and then removed in the end to have a comparison of the orbit in order to check for long-term drifts during the cycle.
In between the variation steps of the quadrupole, the quadrupole was also set to nominal strength in order to check for beam movement due to other effects, which has to be corrected for.
The variation of the quadrupole strength by $\pm\Delta k$ was in the range of \SIrange[range-phrase=--, range-units=single]{5}{10}{\percent}, chosen depending on the current through the quadrupole and the local beta function.
The quadrupole strength variations were done all in a single cycle, as different injection points with a shift of a few tens of \si{\micro\metre} have been observed at COSY. This was done to avoid additional systematic errors. The pattern of the quadrupole strength variation can be seen in figure~\ref{fig:measurement_pattern}.
Considering hysteresis, the quadrupoles were always ramping the same way, first the acceleration ramp and then the small change of an individual quadrupole on flattop. Thus each measurement point always had the same hysteresis curve. Even if the magnetic field of the quadrupole was not exactly changed by $\Delta k$, but by a slightly different amount due to hysteresis, it will always be the same for all measured points. In addition one does not need to have a symmetric measurement for the merit function, as one tries to find its minimum.

\subsection{Optimal position inside the quadrupole}

Each measurement for one quadrupole is characterized by 50 points (i.e. 50 cycles), where the effect of the strength change was measured for different positions in the quadrupole (i.e. steerer settings).
The choice of 50 points was done to have sufficient information for the determination of the optimal position of the beam inside the quadrupole magnet and to be able to finish the measurement in the given time frame (50 cycles taking approximately 1.5 hours).
For each of the 50 points the measurement procedure described beforehand is used and then the merit function (see (\ref{eq:meritFunction})) is calculated.
Some of the 50 measured points had to be discarded due to beam loss, low beam current at corner points and other issues.
The resolution of the BPM reading (\SI{20}{\micro\metre}) is used to compute the error for the merit value of each point.
Note, that the absolute transverse position of the BPM 
with respect to a fixed reference frame is not needed here.
Then a paraboloid is fitted to the data points, with which one can extract the minimum, i.e. optimal position inside the quadrupole.
The error on the minimum for all the fits is of the order of \SI{10}{\micro\metre}.
Since there is no BPM inside the quadrupole, the two BPMs on either side of the quadrupole are used to interpolate into the quadrupole to determine the beam position inside. This interpolation also takes the steerers and their beam deflections into account.

\begin{figure}[!htbp]
\centering
\includegraphics[width = 0.75\textwidth, trim = 5px 25px 40px 45px, clip]{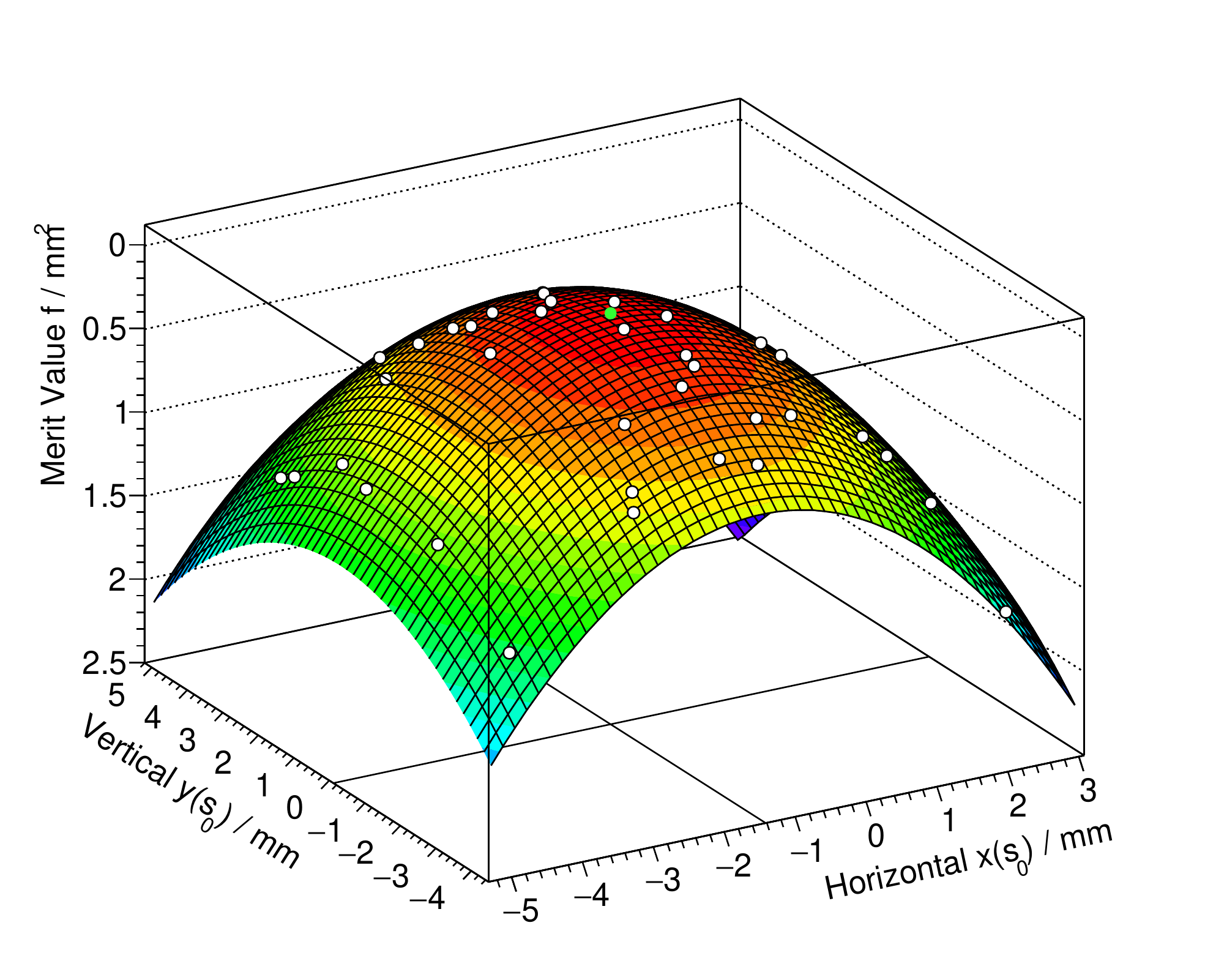}
\includegraphics[width = 0.75\textwidth, trim = 5px 25px 40px 45px, clip]{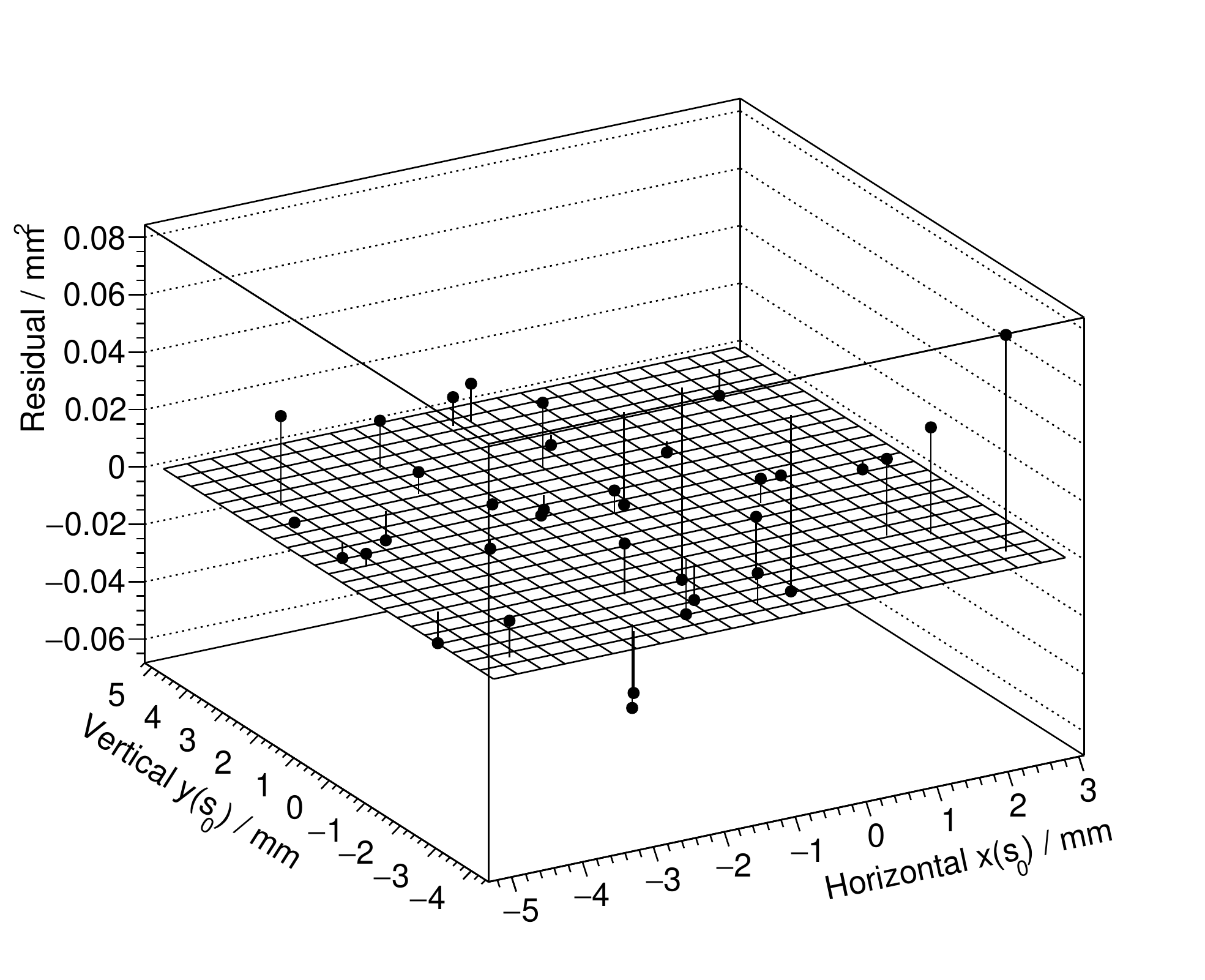}
\caption{An example of a fit for the determination of the optimal position inside a quadrupole is shown. This example shows a  measurement of QU17, which is located in the arcs. The white points are the data points, where on the x- and y-axis the horizontal and vertical displacements of the beam inside the quadrupole are shown, and on the z-axis the calculated merit function $f(x(s_0),y(s_0))$ is depicted. The displacement of the beam inside the quadrupole is obtained by extrapolation from BPMs up- and downstream of the quadrupole. The z-axis has been drawn upside down to make the minimum (highest point in the plot) easier to identify. The data points have a small error ($\approx$\SI{0.008}{\milli\metre\squared}), which is not displayed here. The fit to the data is the colored paraboloid, where the green dot marks the minimum of the fit. In order to guide the eye where the minimum is two lines at the bottom of the plot have been added.
Below the paraboloid plot a residual plot is depicted. It shows the difference of the data points from the fit. Here one can see that for positive horizontal positions there seems to be more of a disagreement between the data and the fit.
}
\label{fig:fitExample}
\end{figure}

An example of such a fit is given in figure~\ref{fig:fitExample}, where one can see the shape of the paraboloid as expected from the merit function.
In addition one can see the optimal position inside that quadrupole as the green point, where the lines at the bottom of the plot are to guide the eye. The quadrupole change $\Delta k$ was kept constant during such a scan.

\begin{figure}
    \centering
    \begin{subfigure}[b]{0.49\textwidth}
        \centering
        \includegraphics[width=\textwidth]{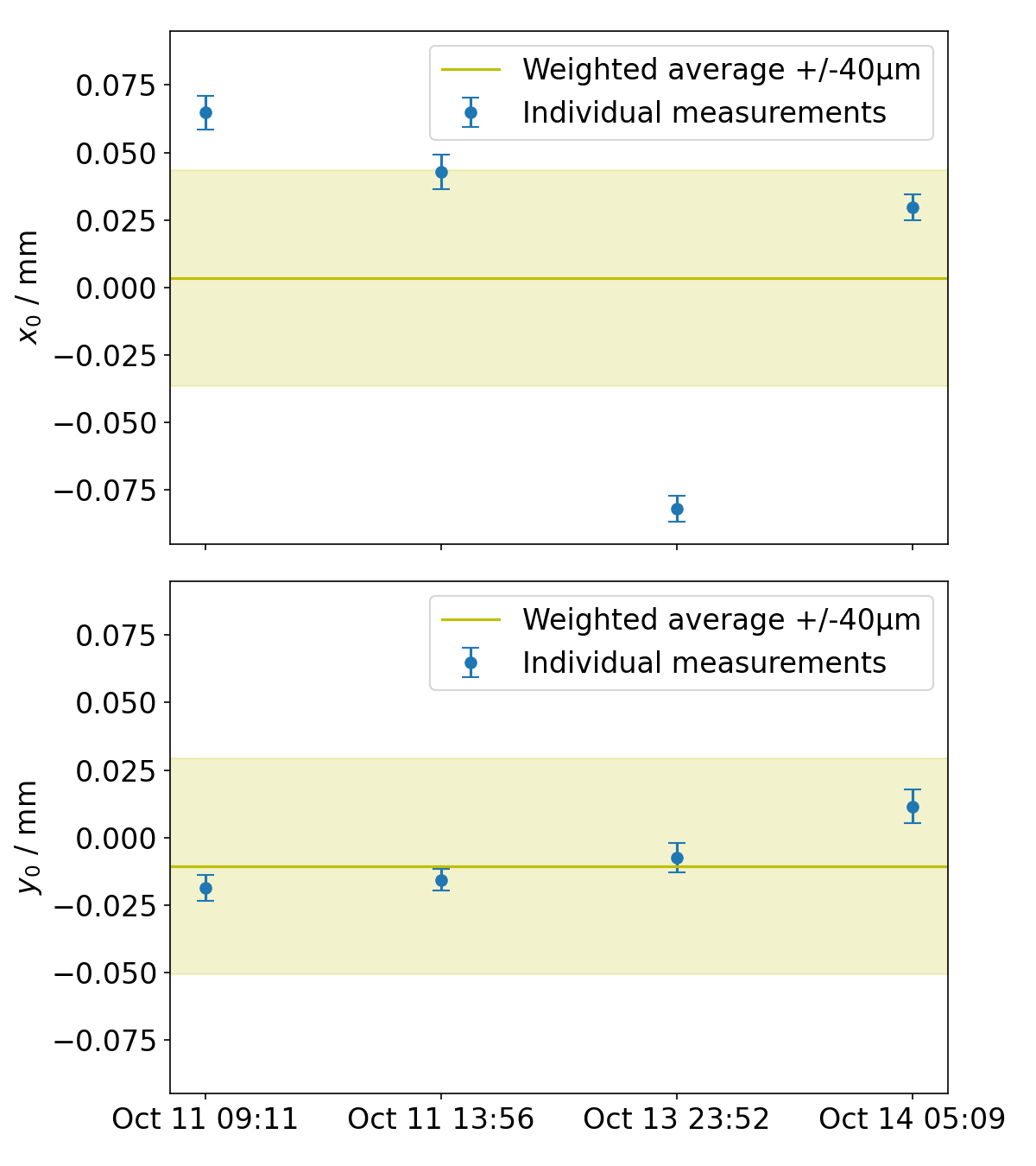}
        \caption{QT04}
        \label{fig:QT04Spread}
    \end{subfigure}
    \begin{subfigure}[b]{0.49\textwidth}
        \centering
        \includegraphics[width=\textwidth]{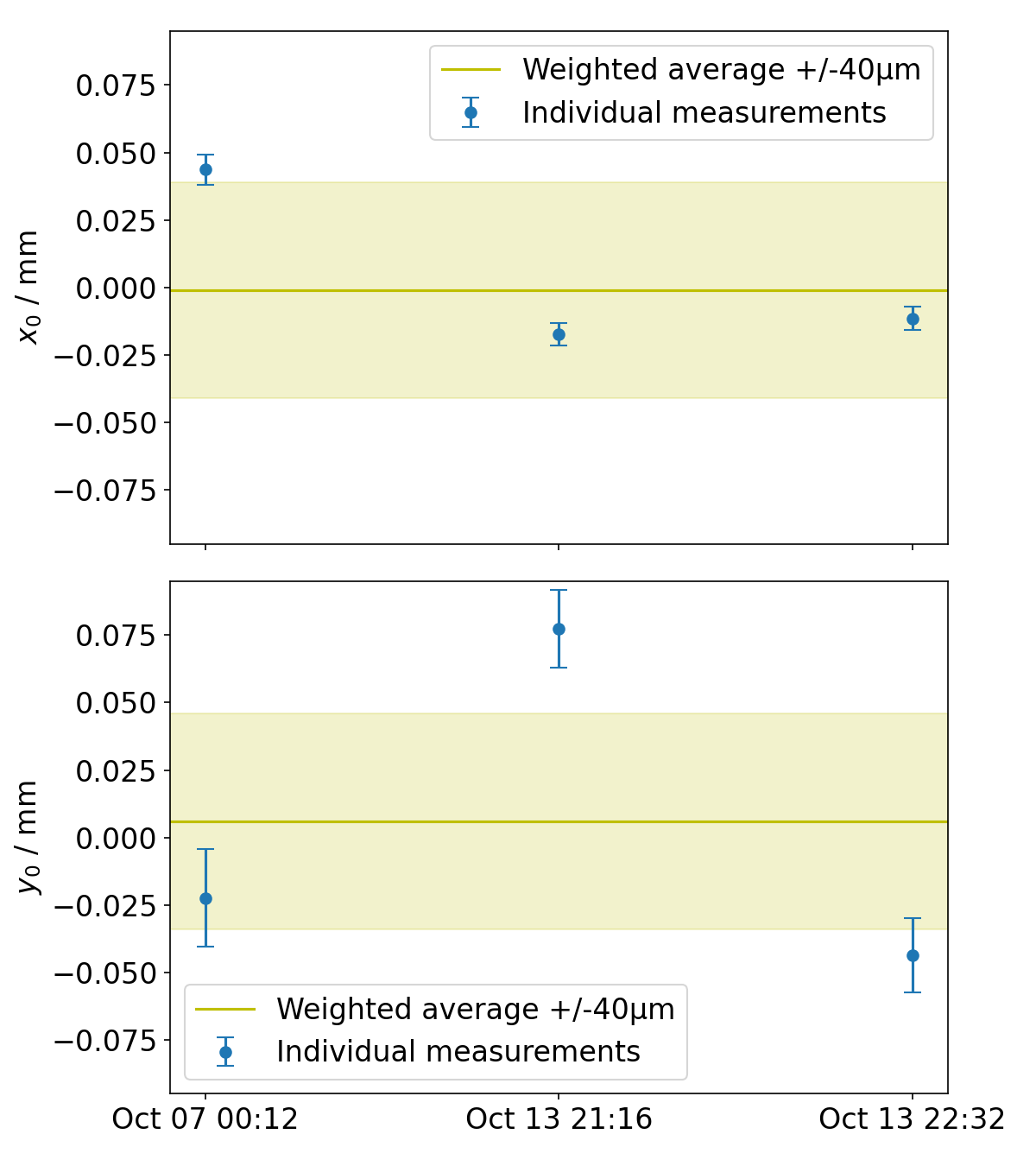}
        \caption{QU11}
        \label{fig:QU11Spread}
    \end{subfigure}
    \caption{Spread of measured optimal positions inside quadrupole QT04 and QU11 respectively. The data points are the results from the individual fits with the errors obtained by the fitting procedure. On the x-axis the date and time of the measurement are shown. In addition the weighted average values of the individual measurements are shown as the horizontal yellow lines. The yellow shaded band around those has a size of $\pm$\SI{40}{\micro\metre}, which is the error assigned to all average positions, as explained in the text.}
    \label{fig:QT28_spread_example}
\end{figure}

The two to four measurements that were taken for each quadrupole were combined to get a value for the optimal position in each of the quadrupoles.  
In some cases the variation of the optimal positions between the individual measurements is around \SI{150}{\micro\metre}, which is larger than the uncertainty on the minimum of the fit ($\approx$\SI{10}{\micro\metre}).
In other cases the variation between individual measurements is nearly zero.
Thus the error on the combined optimal position from the repeated measurements has been estimated by looking at the distribution of the different spreads of all the quadrupoles to get an estimate on how much repeated measurements differ from each other.
From this an error of \SI{40}{\micro\metre} has been calculated and applied to all optimal positions. An example of the spread of the repeated measurements can be seen in figure~\ref{fig:QT28_spread_example}, where the individual measurements and their errors from the fit are displayed. In addition a weighted average of the individual measurements with an error band of \SI{40}{\micro\metre} is shown.
The variation of up to \SI{150}{\micro\metre} can be explained by the movement of the magnets themselves. There was a long-term measurement of two quadrupoles with a laser tracker, which monitored their position over a month. There one could also see a position variation of about \SI{100}{\micro\metre} over the day-night cycle. This effect does not show up for all magnets, as the beam pipe, where the BPMs are mounted to follows the movement to some extent.

\begin{sidewaysfigure}[!htbp]
\centering
\includegraphics[width = \textheight]{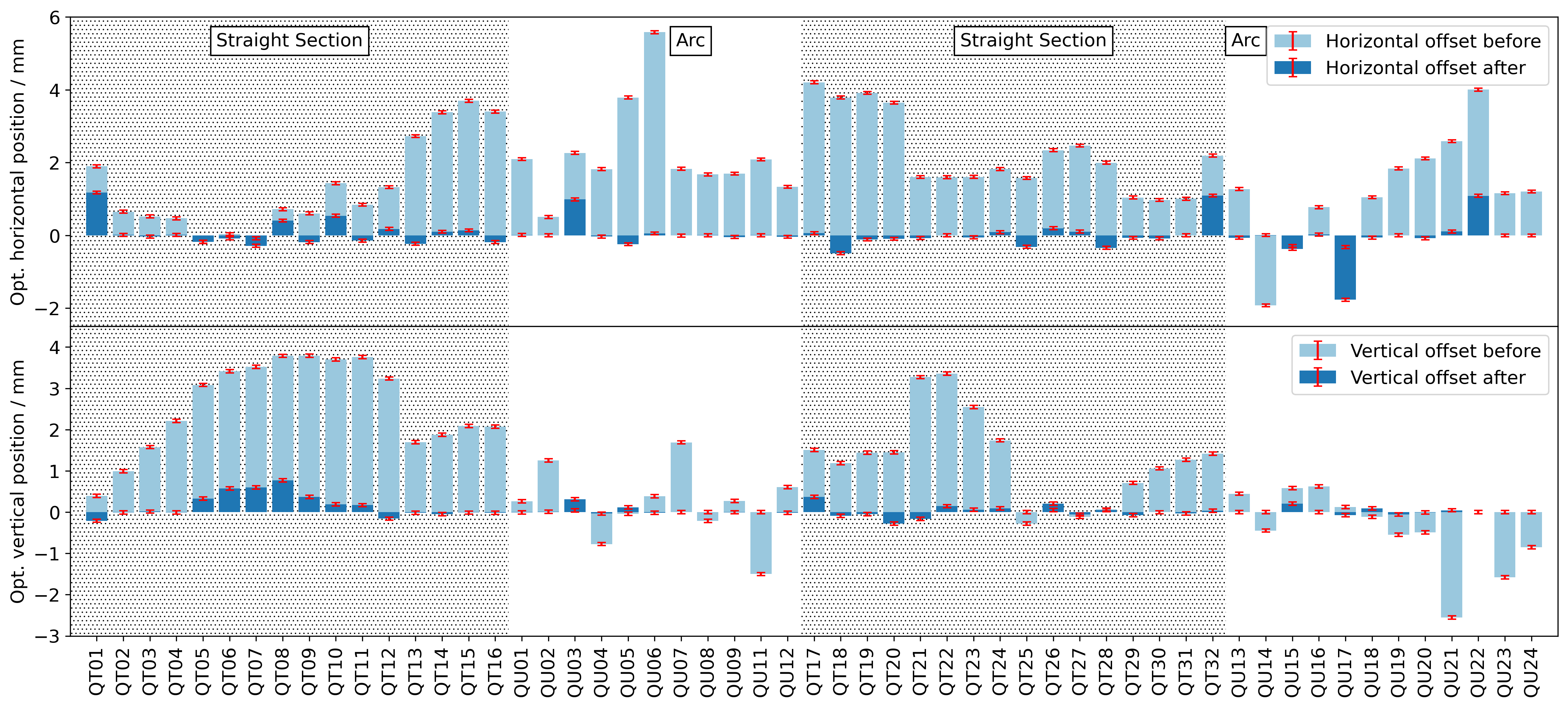}
\caption{The optimal beam position in all quadrupoles. The top part of the plot shows the horizontal direction and the bottom part shows the vertical direction. The optimal position inside the quadrupole before the BPM offset calibration can be seen in the light blue color and after the BPM offset calibration it is shown in the dark blue color. The error on the optimal positions is \SI{40}{\micro\metre} as indicated by the red error bars.
The optimal positions before the offset calibration are not close to zero, which is corrected after the offset calibration, as the optimal positions have been pulled closer to zero.
In the straight sections the quadrupoles labelled QT are close together in sets of four and are expected to be on the same axis, as they refer to the same BPMs.
Thus one can fit a straight line through them to calibrate the BPMs there. For the arcs, where the quadrupoles are labeled with QU, this is not the case, as they are distributed more equally along the arc.
After the offset calibration one can still see some patterns that deviate from zero, which correspond individual quadrupoles that are off by up to \SI{1.2}{\milli\metre}.
In the straight sections one can compare that to the other three quadrupoles in the set and see a misalignment of the quadrupole.
In the arcs the three deviating quadrupoles without a BPM close by and thus one can not pull them to the zero line.
There it is not clear which quadrupole could be misaligned as a comparison is not possible and one has to trust in the mechanical alignment to be correct.
}
\label{fig:QuadrupoleOptimums}
\end{sidewaysfigure}

The resulting optimal position in terms of the uncalibrated offsets of the BPMs in all the quadrupoles are shown in figure~\ref{fig:QuadrupoleOptimums} with the light blue bars. There one can easily see that optimizing the beam to the zero position with the uncalibrated offset will lead to a beam not passing through the center of the quadrupoles.
In addition one can also see that the quadrupoles in the straight sections which are close together (see figure~\ref{fig:COSY_sketch}) are usually on one axis (compare table~\ref{tab:quad_alignment_straight}), which is expected as all quadrupoles are aligned mechanically with a precision of \SI{0.2}{\milli\metre} to the coordinate system of COSY.
This is the case in the straight sections where the quadrupoles are close together in sets of four, whereas in the arcs they are more equally distributed. 

\subsection{BPM offset calibration}

\begin{figure}
    \centering
    \includegraphics[width = 0.7\textwidth]{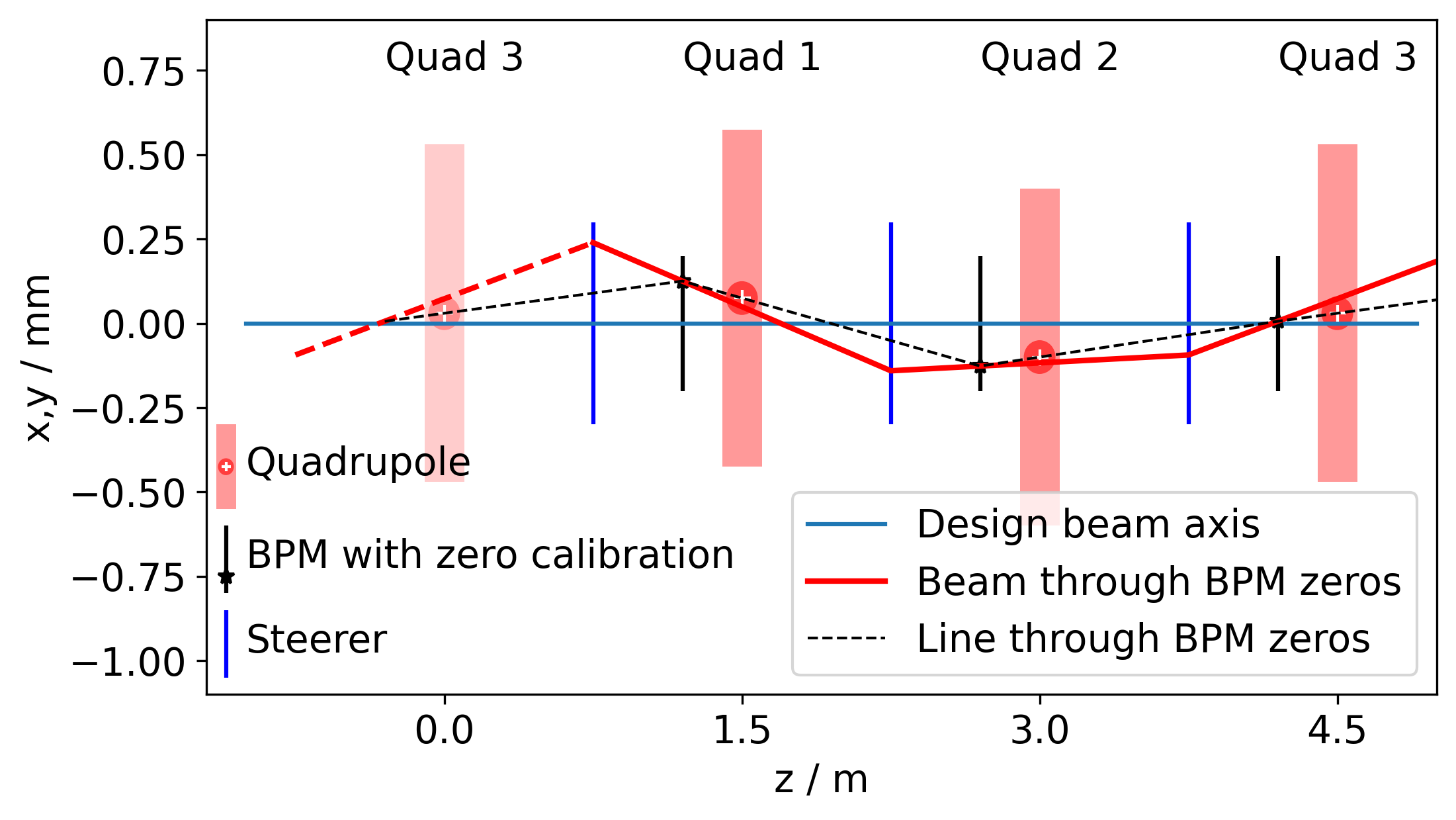}
    \caption{Model sketch of the optimization of the beam path in the accelerator. This model contains only three quadrupoles and has a total ring length of \SI{4.5}{\metre}. The beam-based alignment re-calibrates the BPMs such that the quadrupoles are on the zero line of the coordinate system of the BPMs (indicated by the black dashed line).
    Then, when the beam passes through the zeros of the BPMs it will also pass very close to the quadrupole centers. 
    The optimal beam path does not exactly match the design beam axis, as the quadrupoles are only aligned with an accuracy of \SI{200}{\micro\metre}.
    With respect to the quadrupole centers the beam can be aligned to 
    a precision of \SI{40}{\micro\metre}.
    }
    \label{fig:optimal_orbit_sketch}
\end{figure}

\begin{figure}[!htbp]
\centering
\includegraphics[width = 0.7\textwidth]{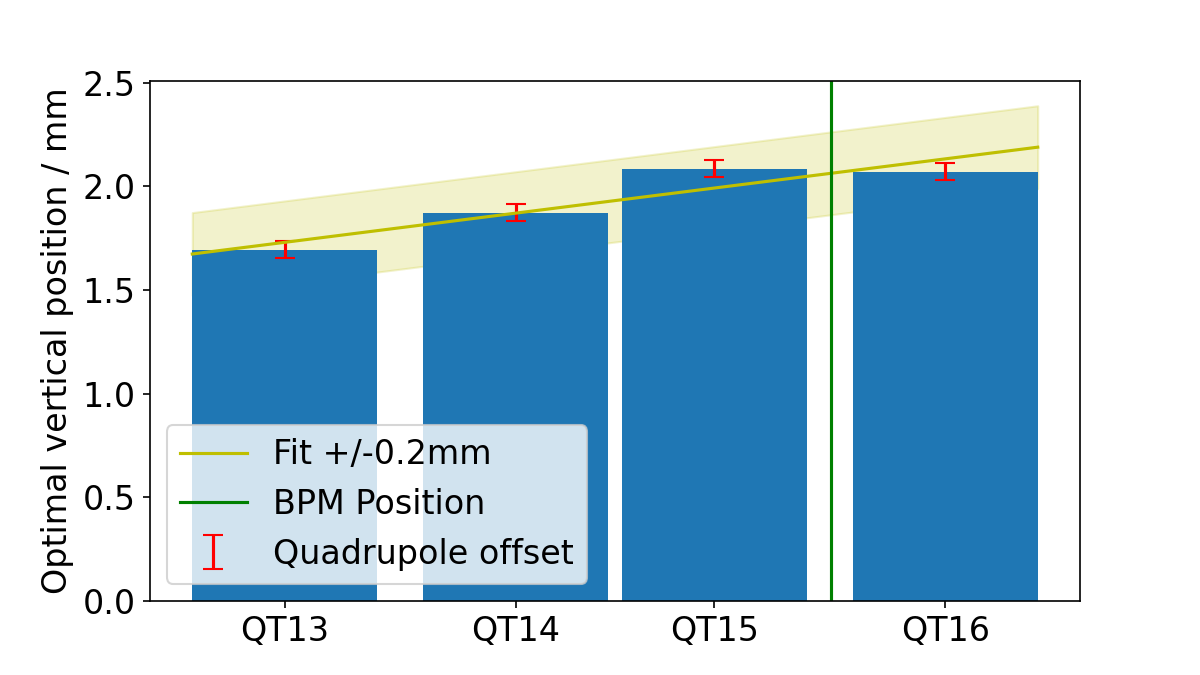}
\includegraphics[width = 0.7\textwidth]{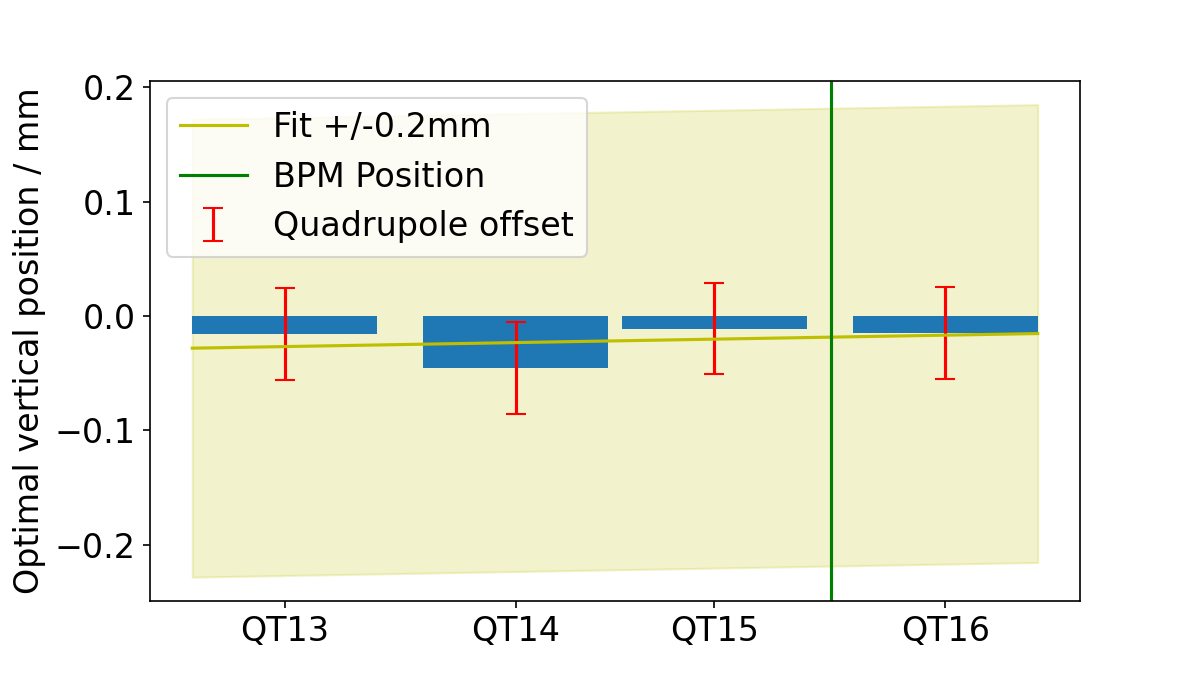}
\caption{In order to calibrate a BPM in the straight sections all four quadrupoles were used to calculate the BPM offset. The bars are the optimal positions in the quadrupoles, where one can fit a straight line (top plot). With that line one can then calculate the offset at the position of the BPM, which is the new BPM offset calibration. The optimal quadrupole position after the BPM offset calibration can be seen in the lower plot, where the optimal quadrupole positions are all close to zero. The shaded region around the fit is used to indicate the alignment precision that the company Stollenwerk achieved. For this specific set of quadrupoles it was better than \SI{0.2}{\milli\metre}, but this is not the case for all of them.
}
\label{fig:exampleBPMCalibration}
\end{figure}

With the now known optimal positions inside the quadrupoles one can calibrate the BPMs such that the new zero in the BPMs corresponds to the quadrupoles also being at the zero of the coordinate system (see black dashed line in figure~\ref{fig:optimal_orbit_sketch}).
The BPM offset calibration is quite straightforward with all the optimal quadrupole positions known.
This will move the optimal position inside the quadrupoles close to the zero orbit in the BPMs, as there are more quadrupoles than BPMs.
In addition one can use the sets of four quadrupoles in the straight sections for the offset calibration as nearly all of them have a BPM inside them and are aligned mechanically.
Thus, not only the two closest quadrupoles but the whole set is used for the offset calibration.
An example for the offset calibration of a BPM can be seen in figure~\ref{fig:exampleBPMCalibration}, where the offset was computed with the help of nearby quadrupoles.

Some observations also resulted from the calculation of the offset calibration, which is that some of the quadrupoles are actually not aligned correctly within the set of four quadrupoles. This will be discussed in more detail later.
In addition a part of the positions inside the quadrupoles could not be moved close to the zero line with the offset calibration of the BPMs, which is also due to a lack of BPMs that can be calibrated in that section, as there are more quadrupoles than BPMs, see figure~\ref{fig:QuadrupoleOptimums}.

\section{Discussion of results}\label{sec:res}

\subsection{Optimal position inside the quadrupoles}

For each quadrupole the optimal position has been extracted from the measured data as described above. These positions then have been used to calibrate the BPMs properly in order to have the zero orbit (see black dashed line in figure~\ref{fig:optimal_orbit_sketch}) in the center of the quadrupoles.
With the new offset calibration of the BPMs one can recalculate the optimal positions in the quadrupoles in that coordinate system and see the improvement, that the centers of the quadrupoles are now at or close to the zero line of the coordinate system.
This can be seen in figure~\ref{fig:QuadrupoleOptimums}, where one can compare the light blue bars, which are the calculated optimal positions inside the quadrupoles before the offset calibration and the dark blue bars, which are after the offset calibration.

\subsection{Alignment of the quadrupoles}
As mentioned before the procedure requires that all quadrupoles are mechanically aligned. According to the surveying this is the case within a tolerance of \SI{0.2}{\milli\metre}.
What one can also see in figure~\ref{fig:QuadrupoleOptimums} is that not all optimal positions in the quadrupoles could be moved close to the zero line. For the straight sections, where there are sets of quadrupoles close together, one can check if individual quadrupoles are not correctly aligned, which is the case for some of them.
This can be seen for example for quadrupole QT01, which is not on the same axis as the rest of the set (see figure~\ref{fig:QuadrupoleOptimums}, where the first four optimal positions inside the quadrupoles do not fit on one line for the horizontal direction).
This observation has been further investigated with a local re-measurement of the mechanical alignment of the quadrupoles, where such an effect can be seen. The mechanical shift of the magnetic center has been verified by observing a small rotation of the quadrupole, which leads to a shift of the magnetic center off of the beam axis like observed with the beam-based alignment.
For the arc sections of the accelerator this comparison is not possible, as there are multiple elements in between the individual quadrupoles. The outliers in the arcs are the quadrupoles, which do not have a BPM close by and thus could not be perfectly accounted for. Here one has to trust the mechanical alignment and assume that all the quadrupoles in the arcs are correctly positioned.

A further observation in figure~\ref{fig:QuadrupoleOptimums} is that there is a pattern in one part of the straight section (QT04-QT12, vertically), where the optimal position inside the quadrupoles rises and then falls. This effect could also not be corrected for with the offset calibration of the BPMs due to technical reasons.

\subsection{BPM offset calibration}

\begin{figure*}[!htbp]
\includegraphics[width = \textwidth, trim = 0px 0px 0px 0px, clip]{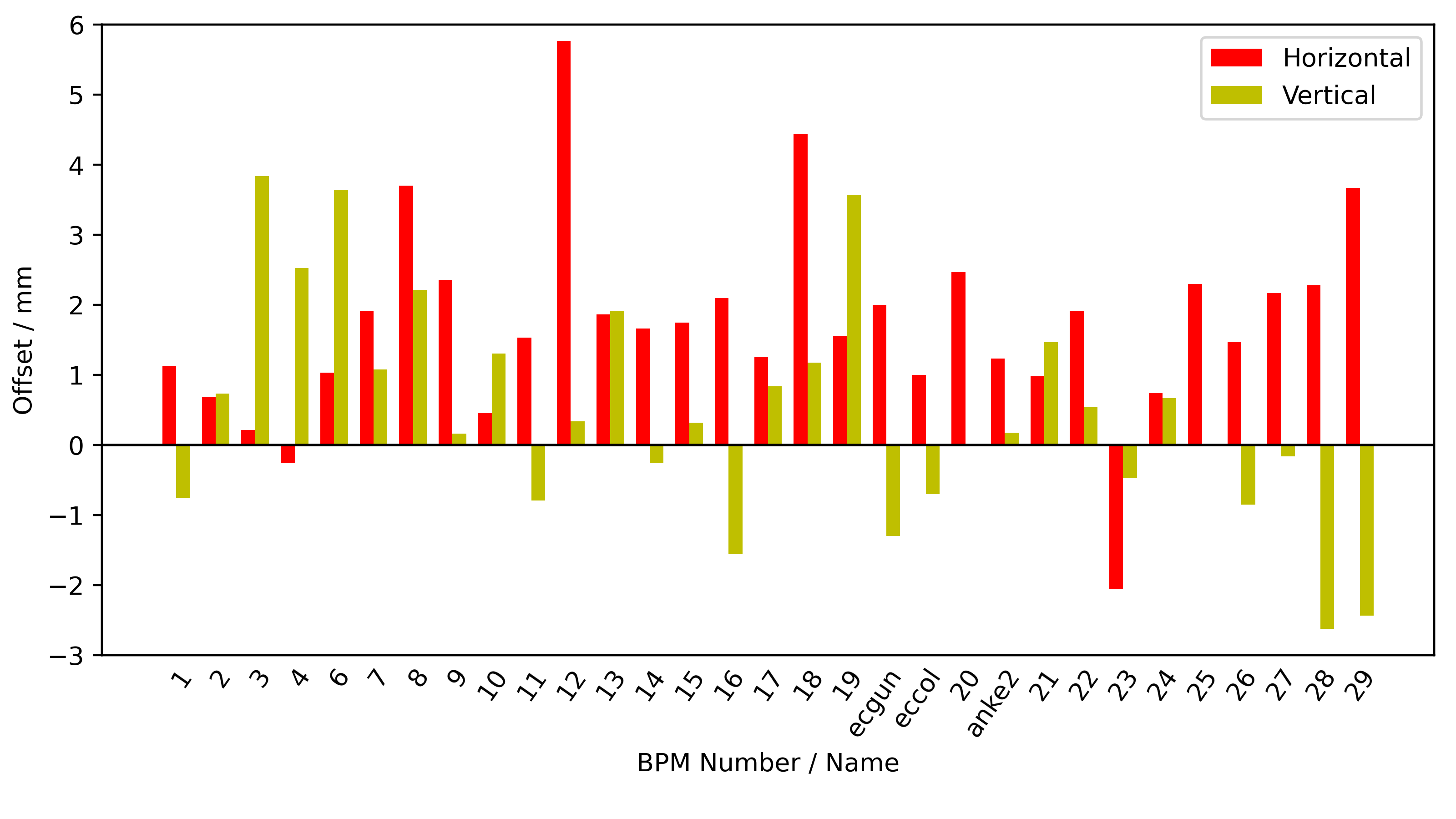}
\caption{The new BPM offset calibration. The horizontal offsets are shown in red and the vertical ones in yellow. On the x-axis the BPM name is displayed and on the y-axis the corresponding offsets. Before the beam-based alignment was done most of the offsets were zero and the BPMs were not properly calibrated. One sees that the BPMs are off by several \si{\milli\metre} with respect to the optimal beam axis given by the magnets.}
\label{fig:newBPMCalibration}
\end{figure*}

The offset calibration of the BPMs has been calculated as explained above and is depicted in figure~\ref{fig:newBPMCalibration}. There one can see the offset calibration that has to be applied to the BPMs in total.
Not included in the bars are mechanical shifts of BPMs, which have been introduced on purpose, e.g. BPM No. 25, which is a BPM close to the extraction.

The overall pattern in the offsets is that for the vertical direction the offset tends to be positive. This can be explained by the fact that the BPMs are mounted on the beam pipe, which itself is mounted on some fixed points, but otherwise lying in the iron yokes of the magnets.
Without further support this causes a shift downwards, thus a positive offset has to be applied.
For the horizontal direction also a trend towards positive values can be seen as well, but here no easy explanation is obvious.
All offsets were applied to the BPMs for future experiments at COSY.

\subsection{Improvement of the orbit}

Now the orbit in the accelerator will improve, but some steering power is still needed, as the mechanical alignment of the quadrupoles is only \SI{200}{\micro\metre}, whereas we could determine the optimal positions inside them with a precision of \SI{40}{\micro\metre}. Thus, the design beam axis of the accelerator (blue horizontal line in figure~\ref{fig:optimal_orbit_sketch}) will not exactly match the optimized orbit in the machine. 
The optimized orbit, going through the zero reading of the calibrated BPMs (red line in figure~\ref{fig:optimal_orbit_sketch}), will be significantly closer to the center of the quadrupoles than the design orbit, as the quadrupoles are slightly off the design axis due to their alignment precision.
The usual algorithm to optimize the orbit in a storage ring is to minimize the BPM reading after the offset calibration was performed. In the ideal case, as depicted in figure~\ref{fig:optimal_orbit_sketch}, all BPM readings are zero. Please note, that this does not necessarily mean that the orbit passes through the center of the quadrupoles as can also be seen in figure~\ref{fig:optimal_orbit_sketch}.
In principle one could use an algorithm which steers the beam through the quadrupole centers (and thus not through the newly calibrated offsets of the BPMs).
Such an algorithm is of course more difficult to implement since it requires particle tracking through quadrupoles as an input. For a high precision experiment like the measurement of electric dipole moments of charged particles~\cite{Abusaif:2019gry}
one tries to avoid additional magnetic fields experienced by the particles because it leads to unwanted spin precessions. An orbit through the quadrupole centers is thus desirable and would usually define the golden orbit.
But this may also require additional steerer power which adds more unwanted magnetic fields.
For the measurement of EDMs the optimization goal could thus even be to find a setting which minimizes the additional magnetic fields seen by the reference particles in steerers and quadrupoles combined.
Given the fact that the accuracy of our beam-based alignment measurements (\SI{10}{\micro \metre} for a single measurement) is much smaller than differences of measurements taken a few days apart (up to \SI{150}{\micro \metre}) show that for a high precision storage ring for electric dipole moment measurements of charged particles one may even need an active orbit feedback system to respond to drifts. This is in addition to a accelerator design that avoids drifts of \SI{150}{\micro \metre} in the first place.

In order to judge by how much the orbit improved due to the correct calibration of the BPM offsets, two measurements were performed. One with the BPM calibration before applying the offsets and the other one afterwards.
For both measurements we tried to correct the orbit as good as possible with the orbit correction software~\cite{IKP2016} used at COSY.
It tries to move the orbit as close as possible to the predefined golden orbit, which corresponds to zero readings in the BPMs.
This is done by using the steerers in the accelerator.

For the first measurement, before applying the offsets of the BPMs, the resulting steerer current RMS can be seen in table~\ref{tab:RMS}.
The RMS values for the second measurement with the applied offsets from the beam-based alignment procedure can also be seen in table~\ref{tab:RMS}. 
In order to see the improvement one has to achieve similar orbit RMS values, which was the case for these measurements, as the orbit correction software ran until the best solution was found. Then a comparison of the steerer current tells by how much the new offset calibration is an improvement.
For the horizontal direction one needs 20\% less steerer current and for the vertical direction 80\% less steerer current, while keeping similar orbit RMS values for both directions.
This improvement shows that one does not have to correct against the beam being offset inside the quadrupoles anymore and thus the beam is not deflected by the quadrupoles anymore.
Unfortunately there were constraints limiting the performance of the orbit correction in the horizontal direction at the time of the measurement. These constraints were, that four steerers in a straight section in COSY could, at the time of the test, not be varied. As these constraints were present in both measurements of the comparison, with and without the new BPM offset calibration, it can still be used to show an improvement.

This result shows that the beam-based alignment has been successfully applied at COSY and it helps improve the orbit in the accelerator. This then also enables a comparison of the measurement with the simulations, as the BPMs are now calibrated, and also to compare simulations to previous measurements.

\begin{table}[!htbp]
\caption{
Change of Steerer current RMS depending on the calibration of the BPM offsets with similar corrected orbits RMS.
Before the offset calibration with the obtained results only known and deliberate shifts of BPMs were included, which was the case for 3 out of 31 BPMs.
After the offset calibration all of the BPMs were calibrated to show zero when the beam is centered in the nearby quadrupoles.
The orbit was corrected to minimal orbit RMS, where the goal was to reach a zero orbit, and the values for the corresponding steerer currents, which are given in a percentage of the maximal current and their corresponding kick in \si{\milli\radian}, were recorded.
Due to constraints during this test the performance of the horizontal direction was not as good as it could have been.
The constraints that were limiting the performance in the horizontal direction, but did not limit the performance in the vertical direction, were, that four steerers in a straight section in COSY could, at the time of the test of the performance, not be varied. As these constraints were present in both measurements of the comparison, with and without the new BPM offset calibration, it can still be used to show an improvement.
}
\label{tab:RMS}

\centering
\bgroup
\def\arraystretch{1.15}
\begin{tabular}{cccc}
\toprule
\multicolumn{2}{c}{Horizontal Steerer RMS$_x$} \\\midrule
Before calibration & \num{5.03}\,\% / \SI{0.63}{\milli\radian} \\
After calibration & \num{3.90}\,\% / \SI{0.49}{\milli\radian}\\
\midrule
\multicolumn{2}{c}{Vertical Steerer RMS$_y$} \\\midrule
Before calibration & \num{4.39}\,\% / \SI{0.25}{\milli\radian}\\
After calibration & \num{0.79}\,\% / \SI{0.05}{\milli\radian}\\
\bottomrule
\end{tabular}
\egroup
\end{table}

\section{Conclusion}

With the beam-based alignment procedure we succeeded in aligning the beam with respect to the center of the quadrupoles to \SI{40}{\micro\metre}.
This is an important ingredient for spin tracking based on a further improved COSY model, in order to finally be able to understand systematic errors of the EDM measurement at COSY \cite{Poncza:2019ith}.
The beam position monitors (BPMs) were calibrated such that the quadrupoles are located at (or close to) the zero line of the coordinate system defined by the BPMs.
The quadrupoles themselves are aligned to a precision of \SI{200}{\micro\metre} with respect to the design beam axis, see figure~\ref{fig:optimal_orbit_sketch}. 

In principle the method could be further improved.
The limit of \SI{40}{\micro\metre} originates from fluctuations between measurements with some time gap in between (see figure~\ref{fig:QT28_spread_example}) where mechanical drifts of this order, due to e.g. temperature changes, are expected.
A single measurement reaches an accuracy of about $\approx$\SI{10}{\micro\metre} (figure~\ref{fig:QT28_spread_example}).
Thus, running a feedback system and continuously monitoring the quadrupoles one could reach the precision of a single measurement.

As a result of this BPM offset calibration, the orbit correction now leads to an orbit passing close to the center of the quadrupoles.
This could be confirmed by the fact that after the beam-based alignment procedure less steerer correction power is needed to reach the optimal orbit, as one does not have to act against the steering of off-center quadrupoles.

Apart from a better orbit in the machine, also misalignments of quadrupoles were observed and confirmed with a mechanical measurement.
Those observed quadrupole misalignments will be corrected in the future improve the quality of the accelerator further.

\acknowledgments
The authors would like to thank the staff of COSY and especially the power supply group of COSY for providing excellent working conditions and for their support concerning the technical aspects of the experiment. 
This work has been financially supported by an ERC Advanced-Grant (srEDM~\#694340) of the European Union and by the EU Horizon 2020 research and innovation programme, STRONG-2020 project, under grant agreement No 824093. The contribution to the work by N.N.~Nikolaev was supported by the Russian State Program 0033-2019-0005.

\appendix

\section{Mechanical alignment of the quadrupoles}
The appendix contains two tables listing the values for the mechanical alignment of the quadrupoles. The goal for the mechanical alignment of the quadrupoles was to align them better than \SI{0.2}{\milli\metre} in transverse (x) direction, better than \SI{0.5}{\milli\metre} in height (y) and better than \SI{1}{\milli\metre} in beam (z) direction. This alignment goal was achieved for most of the quadrupoles. The quadrupoles, which are significantly off in beam direction, e.g. QU23, could not be adjusted further due to nearby installations of other accelerator equipment. The alignment survey by Vermessungsbüro Stollenwerk \& Burghof has been first started in 2016 and in the following years remeasured and readjusted as needed. The data shown in the tables are from 2019.

The measurement is based on measurement-frames, which are mounted on top of markers on the quadrupoles. When those markers were mounted they were measured relative to the field of the quadrupole, thus the alignment of the magnets is the alignment of the magnetic centers and not the mechanical centers of the quadrupoles.

\begin{table}[!htbp]
    \caption{Mechanical alignment of COSY quadrupoles in the straight sections relative to design specifications. $\Delta$z is along beam direction and $\Delta$x and $\Delta$y are horizontally and vertically, respectively. The mean error on those measurements is \SI{0.06}{\milli\metre}. The additional separation in the table indicates the sets of quadrupoles which are located close together. The alignment and measurement of the data has been performed by Vermessungsbüro Stollenwerk \& Burghof.}
    \centering
    \begin{tabular}{lrrr}
    \toprule
        Element & \multicolumn{3}{c}{Translation [mm]}\\ \
         & $\Delta$z & $\Delta$x & $\Delta$y \\ \midrule
        QT01 & -1.21 & 0.02 & -0.37 \\
        QT02 & 0.42 & 0.01 & -0.11 \\
        QT03 & -0.46 & 0.02 & -0.18 \\
        QT04 & 3.43 & 0.07 & -0.34 \\\midrule
        QT05 & 0.39 & -0.04 & -0.06 \\
        QT06 & -0.73 & -0.06 & -0.07 \\
        QT07 & -0.14 & -0.07 & -0.03 \\
        QT08 & -0.62 & -0.08 & 0.98 \\\midrule
        QT09 & -0.33 & 0.03 & 0.06 \\
        QT10 & -0.13 & -0.19 & 0.15 \\
        QT11 & -0.43 & -0.07 & -0.10 \\
        QT12 & -0.45 & -0.03 & 0.09 \\\midrule
        QT13 & -0.34 & 0.08 & 0.21 \\
        QT14 & -0.07 & -0.18 & 0.18 \\
        QT15 & -0.25 & -0.22 & 0.16 \\
        QT16 & -0.33 & -0.09 & 0.01 \\\midrule
        QT17 & 0.11 & -0.13 & 0.56 \\
        QT18 & 0.08 & -0.26 & -0.28 \\
        QT19 & 0.13 & -0.12 & 0.34 \\
        QT20 & -0.92 & -0.23 & 0.24 \\\midrule
        QT21 & 2.72 & -0.31 & 0.35 \\
        QT22 & 0.76 & -0.39 & 0.10 \\
        QT23 & 0.60 & -0.21 & 0.02 \\
        QT24 & 0.75 & -0.27 & 0.12 \\\midrule
        QT25 & 0.45 & -0.28 & 0.04 \\
        QT26 & 0.51 & -0.30 & 0.86 \\
        QT27 & 0.59 & -0.30 & -0.11 \\
        QT28 & 0.70 & -0.19 & -0.04 \\\midrule
        QT29 & 1.78 & -0.16 & -0.12 \\
        QT30 & 0.32 & 0.15 & 0.13 \\
        QT31 & 0.50 & 0.05 & 0.23 \\
        QT32 & 0.43 & -0.24 & 0.16 \\\bottomrule
    \end{tabular}
    \label{tab:quad_alignment_straight}
\end{table}
\begin{table}[!htbp]
    \caption{Mechanical alignment of COSY quadrupoles in the arcs relative to design specifications. $\Delta$z is along beam direction and $\Delta$x and $\Delta$y are horizontally and vertically, respectively. The mean error on those measurements is \SI{0.06}{\milli\metre}. The alignment and measurement of the data has been performed by Vermessungsbüro Stollenwerk \& Burghof.}
    \centering
    \begin{tabular}{lrrr}
       \toprule
        Element & \multicolumn{3}{c}{Translation [mm]}\\ \
         & $\Delta$z & $\Delta$x & $\Delta$y \\ \midrule
        QU01 & -0.69 & -0.14 & -0.10 \\
        QU02 & 0.13 & -0.06 & -0.22 \\
        QU03 & 0.22 & 0.04 & -0.37 \\
        QU04 & 0.68 & 0.04 & -0.40 \\
        QU05 & -0.20 & -0.02 & -0.39 \\
        QU06 & -0.91 & -0.07 & -0.37 \\
        QU07 & -0.12 & 0.12 & -0.24 \\
        QU08 & 0.06 & 0.23 & -0.93 \\
        QU09 & -0.18 & 0.21 & -0.25 \\
        QU10 & -5.22 & 0.25 & -0.38 \\
        QU11 & 0.29 & -0.15 & -0.06 \\
        QU12 & 0.43 & -0.22 & 0.10 \\
        QU13 & -0.18 & -0.05 & 0.11 \\
        QU14 & 0.47 & -0.14 & -0.13 \\
        QU15 & 0.08 & -0.02 & -0.15 \\
        QU16 & -0.12 & -0.02 & -0.06 \\
        QU17 & 0.06 & -0.02 & -0.07 \\
        QU18 & 0.03 & -0.04 & -0.17 \\
        QU19 & 0.28 & 0.14 & -0.20 \\
        QU20 & 0.49 & 0.11 & -0.26 \\
        QU21 & 0.32 & 0.09 & -0.46 \\
        QU22 & -9.78 & 0.11 & -0.47 \\
        QU23 & 16.82 & -0.22 & -0.32 \\
        QU24 & -0.95 & -0.07 & -0.27 \\ \bottomrule
    \end{tabular}
    \label{tab:quad_alignment_arc}
\end{table}

\clearpage
\bibliography{refs}

\end{document}